\documentclass[pre,superscriptaddress,showpacs,nofootinbib,floatfix,preprint]{revtex4}
\usepackage{graphicx}
\usepackage{dcolumn}
\usepackage{bm}
\usepackage{epsfig}
\begin{document}

\preprint{PREPRINT}

\date{\today}

\title{Liquid Polymorphism and Double Criticality in a Lattice Gas Model}

\author{Vera B. Henriques}\email{vhenriques@if.usp.br}
\affiliation{Instituto de F\'{\i}sica, Universidade de S\~ao Paulo,
Caixa Postal 66318, 05315970, S\~ao Paulo, SP, Brazil}
\author{Nara Guisoni}\email{nara@if.usp.br}
\affiliation{Instituto de F\'{\i}sica, Universidade de S\~ao Paulo,
Caixa Postal 66318, 05315970, S\~ao Paulo, SP, Brazil}
\author{Marco Aur\'elio Barbosa}
\affiliation{Instituto de F\'{\i}sica, Universidade de S\~ao Paulo,
Caixa Postal 66318, 05315970, S\~ao Paulo, SP, Brazil}
\author{Marcelo Thielo}\affiliation{Instituto de
  F\'{\i}sica, UFRGS, 91501-970, Porto Alegre, RS, Brazil}
\author{Marcia C. Barbosa}\email{barbosa@if.ufrgs.br}\affiliation{Instituto de
  F\'{\i}sica, UFRGS, 91501-970, Porto Alegre, RS, Brazil}

\date{\today}
\begin{abstract}

We analyze the possible phase diagrams of a simple model for an associating
liquid proposed previously. Our two-dimensional lattice model combines oreintational
ice-like interactions and \"{ }Van der Waals\"{ } interactions which may be
repulsive, and in this case represent a penalty for distortion of hydrogen bonds
in the presence of extra molecules. These interactions can be interpreted in
terms of two competing distances, but not necessarily soft-core. We present mean-field
calculations and an exhaustive simulation study for different parameters which
represent relative strength of the bonding interaction to the energy penalty
for its distortion. As this ratio decreases, a smooth disappearance of the double
criticality occurs. Possible connections to liquid-liquid transitions of molecular
liquids are suggested.

\end{abstract}


\maketitle


\section{\label{sec1}Introduction}


The recently acknowledged possibility of the
existence of single component systems which 
display coexistence between two different
liquid phases \cite{Po92}\cite{Gl99} has opened many interesting questions
regarding the link between the interaction
potential and the presence of double criticality .

Network-forming fluids are primary candidates for 
liquid-liquid transitions. They exhibit 
directional, intermolecular attractions that  lead to  the
formation of bonds between molecules. As a result, locally
structured  regions have lower density than unbonded regions.
These structures at the fluid  phase are transient and 
local but become permanent at lower temperatures.

The case of water is probably the network-forming liquid
 most intensively studied, due to
its ubiquity in nature. An unstable liquid-liquid transition ending in a
critical
 point was initially proposed to explain the anomalous behavior of
network-forming liquids such as water \cite{Po92}\cite{De98}-\cite{An72}.
Closely associated with the possibility of a liquid-liquid phase transition in supercooled water is the phenomenon of polyamorphism occurring below the glass transition temperature, $T_g$. Polyamorphism refers to the occurrence of distinct amorphous solid forms. 
There is by now a general consensus that water displays a transition between
two different amorphous states in the supercooled region of
its phase diagram \cite{Mi98}. Experiments carried out in water at $T\approx 130\; K$ show an abrupt change of volume as a function of
pressure which indicates the existence of a first-order transition \cite{Mi84}.
The two amorphous phases of water might be related to two different
liquid phases at  higher temperatures.  At coexistence, these 
two liquid phases determine a first-order transition line ending 
at a critical point.  Simulations and experiments 
predict  that the liquid-liquid transition is in an experimentally
inaccessible 
region of the phase-diagram \cite{Po92}\cite{Ta96}\cite{Mi00}\cite{Fr03}.

Notwithstanding of its confirmation for metastable water, the coexistence of two liquid phases was uncovered as
a possibility for a few both network-forming  and non-bonding liquids.
Computer simulations for realistic models 
for carbon \cite{Gl99}, phosphorus \cite{Mo01}, germanium \cite{Du02},
silica ($SiO_2$)  \cite{An97}\cite{Sa01} \cite{La00}\cite{Hu04} 
 and silicon \cite{An96}\cite{Alex04} suggest the existence
of a first-order transition between two liquid phases. Recent 
experiments for
phosphorus \cite{Ka00}\cite{Mo03} or phosphate compounds \cite{Ku04},
and for carbon \cite{To97}
 confirm these predictions.

Substances that are structurally similar to water, such as Si, Ge, $GeO_2$ and 
silica, can exhibit not only the liquid-liquid transition but also
the other thermodynamic and dynamic anomalies present in water.
Particular attention has been given to silica because its technological 
applications. 
Many intriguing properties of liquid silica and water occur at low
temperatures. Examples include negative thermal expansion coefficients
and polymorphism
for both liquids. Compression  experiments on silica show a non-trivial change in the microscopic structure \cite{Gr84}\cite{He86}. 
This behavior, reproduced by simulations
 \cite{Sa93}-\cite{Ji93}, was interpreted as an
indication of  a first-order 
transition. Inspired by these results, new simulations
for different models for silica were performed \cite{An97}\cite{Sa01}
\cite{La00}, giving  support to the hypothesis 
of a first-order transition
in silica.
However, experiments must sill confirm a discontinuous transition  between these amorphous phases \cite{Gr84}\cite{Wi88}.

In resume, a full understanding of the effects of the number, spatial
orientation, and strength of bonds on the global 
fluid-phase behavior of network fluids is still lacking.

In order to gain some understanding, a number of simple models 
have been proposed. Since the work of Bernal
\cite{Be33},  the water anomalies have been described 
in terms of the the presence of an extensive hydrogen bond
network which persists in the fluid phase \cite{Er02}. In the case of
lattice models, the main strategy has been to associate the hydrogen bond disorder
with bond \cite{Fr03}\cite{Sa96} or site \cite{Sa93b}\cite{Ro96b}
Potts states. In the former case coexistence between two liquid phases may follow
from the presence of an order-disorder transition and a density anomaly is introduced
\emph{ad hoc} by the addition to the free energy of a volume term proportional
to a Potts order parameter. In the second case, it may arise from the competition
between occupational and Potts variables introduced through a dependency of
bond strength on local density states.

A different approach is to represent hydrogen bonds through ice
variables\cite{Hu83}\cite{attard}\cite{Na91}\cite{Gu00}, so successful in
the description of ice \cite{Li67} entropy, for dense systems. In this case, an
order-disorder transition is absent. Recently  a description based also on ice
variables but which allows for a low density ordered structure \cite{He05} was proposed.
 Competition between the filling
up of the lattice and the formation of an open four-bonded orientational
structure
 is naturally introduced in terms of the ice bonding variables and
no \emph{ad
 hoc} introduction of density or bond strength variations is
needed. Our approach
 bares some resemblance to that of some
continuous models\cite{Si98}\cite{Tr99}\cite{Tr02},
which, however, lack
entropy related to hydrogen distribution on bonds. Also, the reduction of
phase-space imposed by the lattice allows construction
 of the full phase
diagram from simulations, not always possible for continuous
 models
\cite{Si98}. In our previous publication, we have shown that 
 this model is
able to exhibit for a convenient set of parameters both
 density anomalies and
the two liquid phases. Here we explore 
 the phase space for various values
of the orientational energy term. We show that by varying the relative strength
of the orientational 
 part, we can go at a fixed temperature from two coexisting liquid phases as
observed in 
 amorphous water to a smooth transition between two amorphous
structures as might be the case of  silica. 
 
 The remainder of the paper goes
as follows. In sec. \ref{sec2} the model 
 is revised, for clarity. In sec. \ref{sec3}  mean-field results
are given. Sec. \ref{sec4} has the 
 simulations results. Conclusions end this
session.

 
\section{\label{sec2}The Model}
\begin{figure}
 \begin{minipage}[b]{0.5\linewidth}
\centering \epsfig{figure=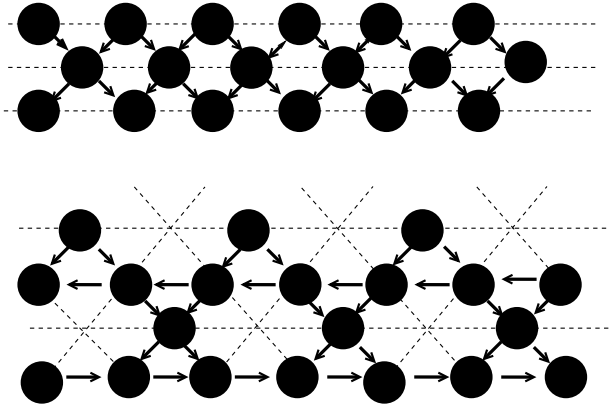,width=\linewidth}
\end{minipage}
\caption{High density liquid, \protect\( HDL\protect \) with density 1
(top) and low density liquid, \protect\( LDL\protect \) with density 3/4
(bottom)on the triangular lattice. The solid lines  indicate the hydrogen
bonds where the arrows differentiate bond donors from bond acceptors.} 
\label{fig1} 
\end{figure}


Consider a lattice gas on a triangular lattice with sites which may
be full or empty. Besides the occupational variables, \( \sigma _{i} \), associated
to each particle \( i \), there are six other variables, \( \tau _{i}^{ij} \),
pointing to neighboring sites \( j \): four are the usual ice bonding arms,
two donor, with \( \tau _{i}^{ij}=1 \), and two acceptor, with \( \tau _{i}^{ij}=-1 \),
while two additional opposite arms are taken as inert (non-bonding), \( \tau _{i}^{ij}=0 \),
as illustrated in Fig. 1. Therefore each occupied site is allowed to be in one
of eighteen possible states.

Two kinds of interactions are considered: isotropic ``van der Waals'' and
orientational hydrogen bonding. An energy \( -v \) is attributed to each pair
of occupied neighboring sites that form a hydrogen bond, while non-bonding pairs
have an energy, \( -v+2u \) (for \( u>0 \)), which makes \( -2u \) the energy
of a hydrogen bond. The overall model energy is given by 
\begin{equation}
\label{E}
E=\sum _{(i,j)}\{(-v+2u)\sigma _{i}\sigma _{j} + u \sigma _{i}\sigma _{j}\tau ^{ij}_{i}\tau ^{ji}_{j}(1-\tau ^{ij}_{i}\tau ^{ji}_{j})]\}
\end{equation}
 where \( \sigma _{i}=0,1 \) are occupation variables and \( \tau _{i}^{ij}=0,\pm 1 \)
represent the arm states described above. Note that each particle may have six
neighbors, but the number of bonds per molecule is limited to four. For \( u/v>1/2 \),
the \"{ }van der Waals\"{ } forces become repulsive. As a result, each molecule
attracts four neighbors, if properly oriented, and repeals the other two. An
interpretation for this \"{ }repulsion\"{ } would be that the presence of the
two extra neighbors distorts the electronic orbitals, thus weakening the hydrogen
bonds. 
\begin{figure}
 \begin{minipage}[b]{0.6\linewidth}
\centering \epsfig{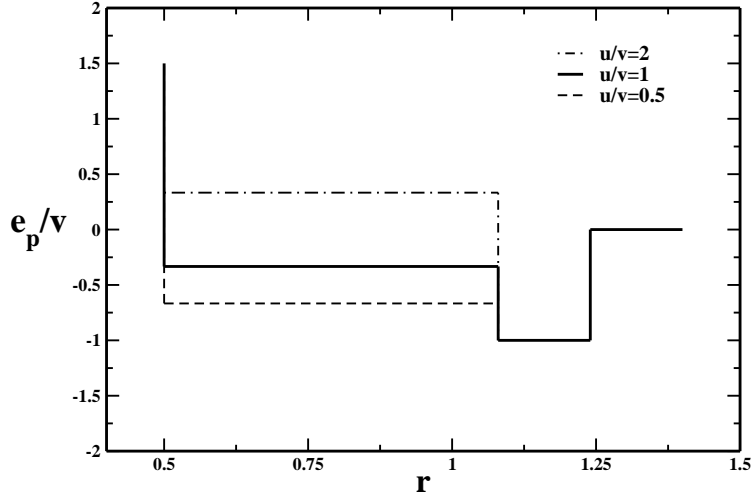}
\end{minipage}
\caption{Effective potential vs inter-particle distance for u/v=0.5 (dashed line),
1 (full-line) and 2 (dot-dashed line).
Penalty on non-bonding neighbors is sufficient to stabilize the low
density phase for u/v=1 and 2. The shoulder
potential is attractive in the first and repulsive in the second case.
For smaller penalty u/v=0.5 only the high density
phase is stable at zero temperature and the step potential is accordingly
"smoother". } 
\label{fig2} \hspace{.8cm}
\end{figure}

Inspection of the model properties allows the prediction of two ordered states,
as shown in Fig. 1. For low chemical potential, the soft core repulsion becomes
dominant, \( \rho =0.75 \), and energy \"{ }volume\"{ } density is given
by \( e=E/V=-3v/2 \), where V is the number of lattice sites. If the chemical
potential is high, \( \rho =1 \), and energy density \( e=-3v+2u \). At zero
temperature, the low density liquid (LDL) coexists with the high density liquid
(HDL) at chemical potential \( \mu /v=-6+8u/v \), obtained by equating the
grand potential density (or pressure) associated with each one of these phases.
Similarly the coexistence pressure at zero temperature is given by \( p/v=-3+6u/v \). 
Besides these two liquid states, a gas phase is also found and it coexists with
the low density liquid at chemical potential \( \mu /v=-2 \) and pressure \( p=0 \).
The condition for the presence of the two liquid phases is therefore \(
u/v>0.5 \), i.e., that the "Van der Waals" interactions are repulsive.

The two ordered structures could be qualitatively associated to different ice
phases. Under pressure, hydrogen bonds reorganize in more dense phases \cite
{eisenberg}.

Our model may be interpreted in terms of some sort of average soft-core potential
for large hydrogen bond energies. The low density  phase implies average interparticle
distance \( \overline{d_{LD}}=\rho _{LD}^{-1/2}=2/\sqrt{3} \), whereas for
the high density  phase we have \( \overline{d_{HD}}=\rho _{HD}^{-1/2}=1 \). The corresponding
energies per pair of particles is \( -v \) and \( -v+2u/3 \). The hard core
is offered by the lattice. For \( u/v>3/2 \),
the shoulder becomes repulsive, making the potential soft-core. Figure
2 illustrates, in a schematic way, the average pair energy $e_p/v$
dependence on distance between 
particles,   for $u/v=0.5,1,2$. Notice that
for $u/v=2$ the potential at high density phases becomes repulsive.
For $u/v=1,0.5$, the shoulder is attractive.


\section{\label{sec3}The Mean-Field Analysis}


Designing a correct mean-field version of the model is not an obvious task,
due to the special orientational ice-like interactions. We have therefore looked
at model properties on the Bethe lattice, for which exact relations may be given.
Each site of the usual Cayley tree (of coordination six) is replaced by a hexagon
and (\( \eta ,\tau  \)) variables are attributed to its vertices (see Fig.
3), with \( \eta  \) and \( \tau  \) defined as in the previous section. This
representation is inspired on the Bethe solution for a generalization of the
square water model \cite{Gu00,square2} presented by Izmailian et al. \cite{izmailian}. For an occupied site-hexagon, we will have two vertices with
\( (\eta ,\tau )=(1,1) \), two with \( (\eta ,\tau )=(1,-1) \) and another
two with \( (\eta ,\tau )=(1,0) \). For an empty site-hexagon the six vertices have \( (\eta ,\tau )=(0,0) \).

\begin{figure}
\begin{minipage}[b]{0.3\linewidth}
\centering \epsfig{figure=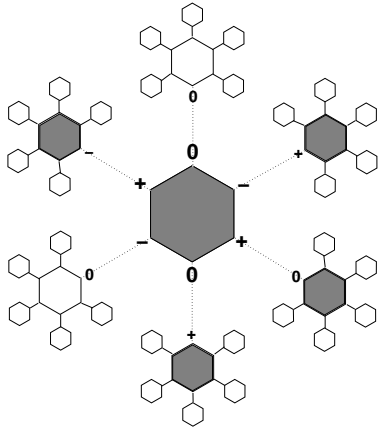,width=\linewidth}
\end{minipage} 
\caption{ The Bernal model on the Bethe lattice. Each site of an usual
Bethe lattice is placed by a hexagon with (\protect\( \eta ,\tau \protect \))
variables placed on the vertices. Sites of the \protect\( N\protect \) generation
are linked by dashed lines.}
\label{fig3}
\end{figure}


The partition function in the grand canonical ensemble is given by: 
\begin{eqnarray}
\Xi (T,\mu )=18e^{\beta \mu }g^{2}_{N}(1,+)g^{2}_{N}(1,-)g^{2}_{N}(1,0)+g^{6}_{N}(0,0), & \label{part} 
\end{eqnarray}
 where \( \mu  \) is the chemical potential and \( g_{N}(1,+) \) is the partial
partition function (generation \( N \)) for a central site with \( (\eta ,\mu )=(1,1) \).
Similarly for \( g_{N}(1,-) \), \( g_{N}(1,0) \) and \( g_{N}(0,0) \). 

On each lattice line we have two \( (\eta ,\tau ) \) pairs of variables connecting
two consecutive generations of the lattice. The Boltzmann weights for each lattice
line are \( \omega =1 \), if at least one of the connecting sites is empty,
and, if both sites are occupied, either \( \omega =e^{\beta v} \), if arm variables
are of opposite sign, and \( \omega =e^{\beta (v-2u)} \), otherwise. 

The density of water molecules at the central site of the tree is given by
\begin{equation}
\label{dens}
\rho (T,\mu )=\frac{18z}{18z+x^{2}_{N}y^{2}_{N}r^{2}_{N}},
\end{equation}
 where \( z=e^{\beta \mu } \) is the activity and 
\[
x_{N}\equiv \frac{g_{N}(0,0)}{g_{N}(1,+)},\hspace {0.3in}y_{N}\equiv \frac{g_{N}(0,0)}{g_{N}(1,-)},\hspace {0.3in}r_{N}=\equiv \frac{g_{N}(0,0)}{g_{N}(1,0)}.\]

Eqn.\ref{dens} gives us a relation between the water density \( \rho  \)
and the activity \( z \): \( 6z=\gamma (x_{N}y_{N}r_{N})^{2} \), with \( \gamma =\rho/(3(1-\rho )) \).
Using this result the recursion relations at the fixed point may be written
as
\begin{eqnarray*}
x=\frac{\gamma (x+y+r)+1}{\gamma e^{\beta (v-2u)}(x+e^{2\beta u}y+r)+1},
\end{eqnarray*}
\begin{equation}
\label{xy}
y=\frac{\gamma (x+y+r)+1}{\gamma e^{\beta (v-2u)}(e^{2\beta u}x+y+r)+1},
\end{equation}
\begin{eqnarray*}
r=\frac{\gamma (x+y+r)+1}{\gamma e^{\beta (v-2u)}(x+y+r)+1}.
\end{eqnarray*}
Under the the physical conditions \( x\geq 0 \), \( y\geq 0 \) and \( r\geq 0 \),
since \( x \), \( y \) and \( r \) are ratios between partition functions,
we must have \( x=y \). Eqs.\ref{xy} reduce to two equations which are
solved numerically for specific values of the model parameters \( u/v \). Chemical
potential vs density isotherms are then obtained from Eq.\ref{dens}, as shown
in Figures  4  and 5.

\begin{figure}
\begin{minipage}[b]{0.55\linewidth}
\centering \epsfig{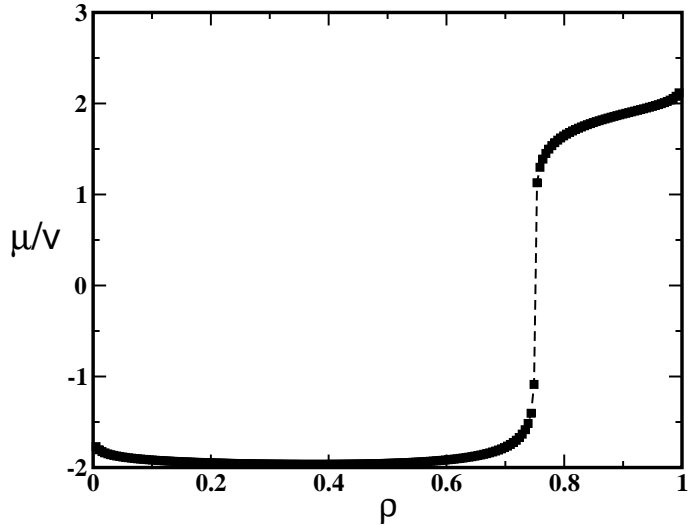}
\end{minipage} 
\caption{ Chemical potential versus water density for 
\protect\( u/v=1\protect \)
and \protect\( \bar{T}=0.05\protect \). The coexistence between 
a gas (\protect\( \rho \sim 0\protect \))
and a low density phase (\protect\( \rho \sim 0.75\protect \)) can be seen
from the van der Waals loop. Note the abrupt increase for 
\protect\( \rho \simeq 0.75\protect \).}
\label{fig4}\hspace{.8cm}
\end{figure}


\begin{figure}
\begin{minipage}[b]{0.55\linewidth}
\centering \epsfig{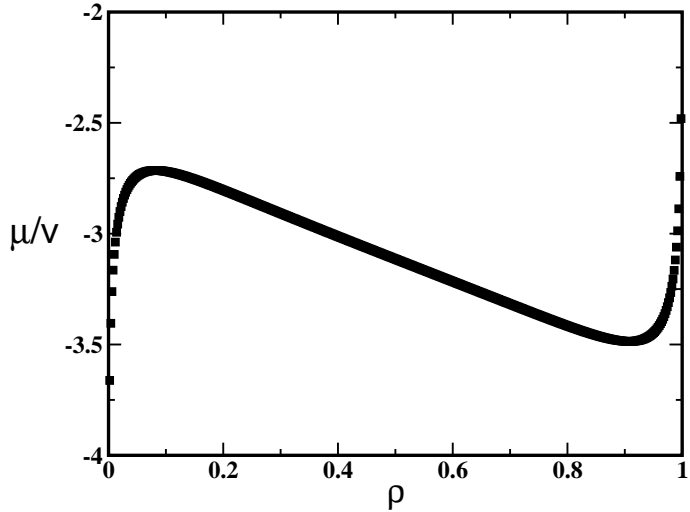}
\end{minipage} 
\caption{Chemical potential versus water density for 
\protect\( u/v=0.25\protect \)
for \protect\( \bar{T}=0.4\protect \). The LD phase is 
no longer present, since coexistence
is between a gas (\protect\( \rho \sim 0\protect \)) and a 
high density phase
(\protect\( \rho \sim 1\protect \)). }
\label{fig5}\hspace{.8cm}
\end{figure}


The first striking result is that temperature destabilizes the liquid-liquid
transition. Van der Waals loops are present only for the gas-liquid, but not
for the liquid-liquid transition. However, for larger bond energies \( u/v>0.5 \),
the liquid phase is of low density \( (\rho =0.75) \), whereas for smaller
bond energies, the liquid phase is of high density \( (\rho =1) \). A typical
isotherm for the first case, for which the gas-liquid transition is to a phase
of low density, is shown in Fig 4, for \( u/v=1 \) . It is to be noted that
the chemical potential presents a very abrupt rise, just above the gas-liquid
transition, as if signaling a quasi liquid-liquid transition. Fig 5 shows a
typical low temperature isotherm for the second case, for \( u/v=0.25 \), for
which no low density phase is found.

Inspite of the absence of a liquid-liquid transition, present in the results
of
 simulations (see \cite{He05} and next section), both the LD and the HD
phases
 appear, for different strengths of the hydrogen bonds.
Also, the abrupt rise of the chemical potential in Fig.4 indicates that this
mean-field treatment comes
near to but misses the LD-HD transition. Similar problems were noted for
bonding ice-type models \cite{izmailian}\cite{square2}.
 Cyclic ordering of edges around a vertex, in the lattice ordered
structure, cannot be represented
on the tree, which may lead to analytical
free-energies for the same parameters for which exact \cite{izmailian} or
simulation \cite{He05} studies predict phase transitions.

\section{\label{sec4}The Monte Carlo Results}

\begin{figure}
\begin{minipage}[b]{0.55\linewidth}
\centering \epsfig{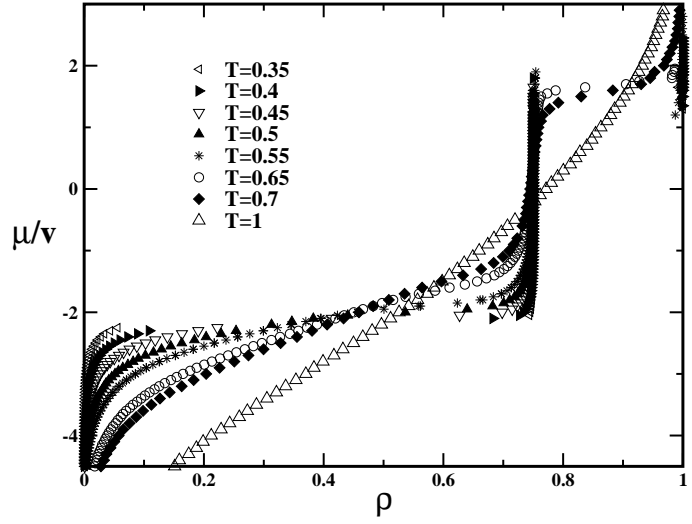}
\end{minipage} \hspace{.2cm}
\caption{Chemical potential vs. density isotherms 
for different temperatures. $\rho$ is given in units of 
lattice space and the $\bar{T}=k_BT/v$. Here we illustrate the case $u/v=1$.}
\label{fig6}\hspace{.8cm}
\end{figure}


\begin{figure}
\begin{minipage}[b]{0.55\linewidth}
\centering \epsfig{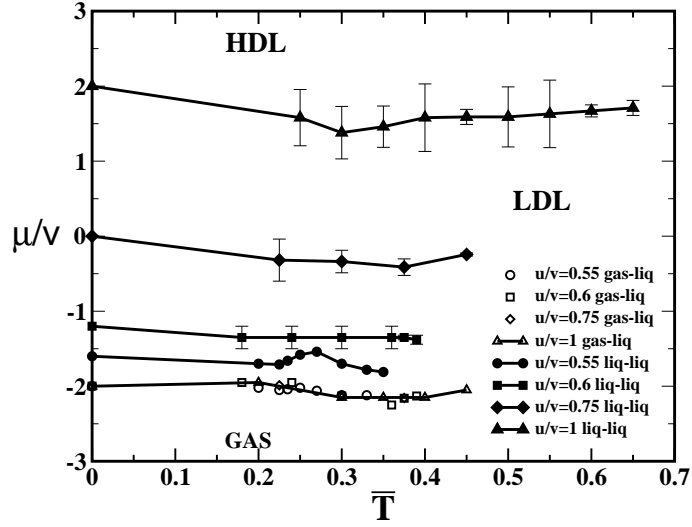}
\end{minipage}
\caption{ Phase-diagram showing chemical potential  vs.  temperature for 
$u/v=0.55,0.6,0.75,1$. The solid lines 
are the  LDL-HDL   coexistence lines. The gas-LDL collapse into 
a single dashed line. 
 The coexistence  at  zero temperature  at $\mu/v=-1.6,-1.2,0,2$  and
at $\mu/v=-2$ are exact. The error bars refer  to the size of the 
hysteresis loop. For visualization purposes the error bars for
the gas-liquid points are not shown.  The statistical error bars 
have similar size as the symbols.
}
\label{fig7}
\end{figure}


\begin{figure}
\begin{minipage}[b]{0.55\linewidth}
\centering \epsfig{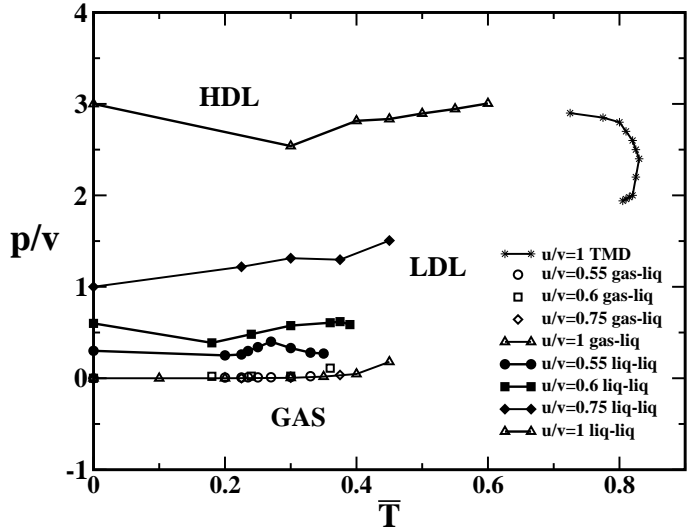}
\end{minipage} 
\caption{ Phase-diagram showing pressure vs.  temperature for $u/v=0.55,0.6,0.75,1$. Reduced pressure $p/v$ is given 
in units of lattice space. The 
solid lines ( and filled symbols)  are  the  LDL-HDL  coexistence lines. The empty symbols
indicate the gas-LDL coexistence lines collapsed into a 
single dashed line. The coexistence 
at  zero temperature  at  $p/v=0.3,0.6,1.5,3$  and $p/v=0$ are exact. 
}
\label{fig8}
\end{figure}


The model properties for finite temperatures were obtained through Monte Carlo
simulations in the grand-canonical ensemble using the Metropolis algorithm.
Particle insertion and exclusion were tested with transition probabilities given
by $w(insertion)=exp(-\Delta \phi )$ and $w(exclusion)=1$  if
$\quad \Delta \phi >0$ or $w(insertion)=1$ and $w(exclusion)=exp(+\Delta \phi)$ if  $\quad \Delta \phi <0$
with \( \Delta \phi \equiv \exp \{\beta (e_{particle}-\mu )-\ln (18)\} \)
where \( e_{particle} \) is the particle energy. Since
the
 empty and full sites are visited randomly, the factor 18 is required in
order
 to guarantee detailed balance.

 A detailed
study of the model properties  
 was presented in a previous publication for a particular value of the
energy parameter, $u/v=1$ \cite{He05}. Here, we present the evolution of the
phase diagram under variation of this energy parameter which represents
relative energy of bond strength. 
 Note that the relevant parameter range is
\( u/v>1/2 \). For 
  \( u<v/2 \), the LDL disappears. 
Our study is for L=10
and runs were of the order of \( 10^{6} \) Monte Carlo steps.

Chemical potential vs. density  
isotherms for 
$u/v=1$,
 and a large set of temperatures are illustrated in Fig.6. The figure 
shows the
  first-order phase transitions between the gas and
the low density liquid phase and  between the low density 
liquid and high density liquid phases.
The coexistence lines are estimated from the mid-gaps
 in the chemical potentials. 
 Fig. 7 illustrates the chemical potential
vs. temperature phase-diagram
 for $u/v=0.55,0.6,0.75,1$. The
gas-LDL coexistence lines  for the different $u/v$
seem to 
 collapse into a single line that at zero temperature
gives 
$\mu/v=-2$.  The different LDL-HDL lines are 
also illustrated. At zero temperature, 
 exact calculations locate the end-points of these lines at, respectively, 
$\mu/v=-1.6, -1.2,0,2$. The error bars here are a measure of 
the size of the hysteresis loops for each temperature. The statistical
errors are of similar size as the symbols.
  As $u/v\rightarrow 0.5$ the liquid-liquid 
coexistence lines  approach the 
gas-liquid transition, merging with it at $u/v=0.5$.

Pressure was 
computed from numerical integration
of the Gibbs Duhem equation, at fixed temperature, from zero pressure at zero
density.
 Fig. 8 shows the corresponding pressure vs temperature phase diagrams.

At zero temperature, we have the exact values $p/v=3,1.5,0.6,0.3$. 
An inversion of the LDL-HDL phase boundary, close to the critical point,
is to be noted. The LDL-HDL line has positive slope for $u/v=1,0.75$, and
negative slope
for $u/v=0.55$.    This indicates that for  $u/v=1,0.75$ 
the  low density
liquid phase is more entropic than the high density 
liquid phase as it would be the case for most liquids.
However, for $u/v=0.55$, the LDL phase is less entropic
than the HDL phase, a behavior that is expected for water.
The line
of temperature of  maximum densities, $TMD$, is present for
all $u/v>0.5$ values. The figure illustrates the $TMD$ only for the $u/v=1$ case.

In summary, we have shown, both from mean-field and from simulations that the
model we proposed in a previous publication may present two liquid phases and
the associated double criticality, depending on the ratio of bond strength to
energy penalty for distortion, represented by our parameter \( u/v \). Double
criticality disappears in case the price paid for increasing the coordination
is too high. A second consequence of varying the relative energies is that the
slope of the liquid-liquid line may become negative, around the critical point. 

Analysis of the model behavior shows that the two liquids are present if the
two competing distances are sufficiently separated, but the \"{ }shoulder\"{ }
need not be negative (see Fig. 2), in contradiction to a recent proposal \cite{Fr01}.
In other words, two characteristic attractive distances might also generate the
liquid-liquid transition.

Experimental and numerical evidence suggests that water possesses a first-order
transition between a low density liquid and a high density liquid. The present
model shows two liquid phases and consequently two critical points and a line
of density anomalies.

Experimental evidence for a first-order transition in silica is still missing.
According to our model, as the bond energy varies, the critical point temperature
decreases. For temperatures above the critical point, under compression the
liquid goes from the low density phase to high density phase in a continuous
way as it is observed experimentally in silica. Since the liquid-liquid critical point moves to lower temperatures as the bond energy becomes weaker, one could suggest that the first-order transition predicted by simulations is not observed
in experiments because the coexistence line and the second critical point in
the case of silica might be located at very low temperatures. 

Water and silica are network bonding systems, and present low-coordination solids
with transition to higher coordinated solids at high pressures. Ice I is tetrahedrally
bonded, with bonds suffering distortion, for higher pressures. At still higher
pressures, the tetrahedral structures interpenetrate, to satisfy both energy
and pressure requirements. Silica transitions between tetrahedrally and octahedrally
coordinated structures. It would be interesting to establish energy criteria
for bond distortions in both cases, in order to test the ideas we propose in
this study.

\vspace*{1.25cm}

\noindent \textbf{\large Acknowledgments}{\large \par}

\vspace*{0.5cm} This work was supported by the Brazilian science agencies CNPq,
FINEP, Fapesp and Fapergs.

\end{document}